\documentclass[prl, reprint, superscriptaddress, amsmath, amssymb, aps]{revtex4-1}

\usepackage{graphicx}   

\usepackage{etoolbox}
\apptocmd{\thebibliography}{\raggedright}{}{}

\newcommand{\vc}[1]{\boldsymbol{#1}}
\newcommand{\Ai}{\mathrm{Ai}}

\begin{document}

\title{Phase memory preserving harmonics from abruptly autofocusing beams}

\author{Anastasios.~D.~Koulouklidis}
\affiliation{Institute of Electronic Structure and Laser (IESL), Foundation for
             Research and Technology - Hellas (FORTH), P.O. Box 1527, GR-71110
             Heraklion, Greece}
\affiliation{Materials Science and Technology Department, University of Crete,
             71003, Heraklion, Greece}

\author{Dimitris~G.~Papazoglou}
\email{dpapa@materials.uoc.gr}
\affiliation{Institute of Electronic Structure and Laser (IESL), Foundation for
             Research and Technology - Hellas (FORTH), P.O. Box 1527, GR-71110
             Heraklion, Greece}
\affiliation{Materials Science and Technology Department, University of Crete,
             71003, Heraklion, Greece}

\author{Vladimir Yu. Fedorov}
\affiliation{Science Program, Texas A\&M University at Qatar, P.O. Box 23874,
             Doha, Qatar}
\affiliation{P.~N.~Lebedev Physical Institute of the Russian Academy of
             Sciences, 53 Leninskiy Prospekt, 119991, Moscow, Russia}

\author{Stelios~Tzortzakis}
\affiliation{Science Program, Texas A\&M University at Qatar, P.O. Box 23874,
	Doha, Qatar}
\affiliation{Institute of Electronic Structure and Laser (IESL), Foundation for
             Research and Technology - Hellas (FORTH), P.O. Box 1527, GR-71110
             Heraklion, Greece}
\affiliation{Materials Science and Technology Department, University of Crete,
             71003, Heraklion, Greece}

\date{\today}

\begin{abstract}
We demonstrate both theoretically and experimentally that the harmonics
from abruptly auto-focusing ring-Airy beams present a surprising property, they
preserve the phase distribution of the fundamental beam.
Consequently, this ``phase memory'' inherits to the harmonics the abrupt
autofocusing behavior, while, under certain conditions, their foci coincide in
space with the one of the fundamental.
Experiments agree well with our theoretical estimates and detailed numerical
calculations.
Our findings open the way for the use of such beams and their harmonics in
strong field science.
\end{abstract}

\maketitle

{\it Introduction.}---%
The preservation of wave packet phase, after the action of nonlinear or other
phase deteriorating effects, is often referred to as ``phase
memory''~\cite{Coskun1999,Basharov2006,Smithey1991}.
Phase memory in general leads to a coherent wave packet
behavior~\cite{Coskun1999} while its physical origin is quite diverse ranging
from quantum-mechanical physical system configurations~\cite{Basharov2006,
Smithey1991}, to optical wave packet distributions~\cite{Coskun1999}.
Optical wave packets that exhibit phase memory present exciting applications in
strong field
physics, like in higher harmonics and attosecond pulses~\cite{Chini2014}.

Optical harmonics are generated by exploiting strong field interactions in
nonlinear optical media.
An intense pulsed beam of fundamental frequency $\omega$ with electric field
amplitude ${E}(\vc{r},t)$  will modulate the dielectric polarization density
$P(\vc{r},t)$ of the medium.
This physical process can be described as~\cite{Boyd2003}:
$P(\vc{r},t) = \varepsilon_0 \chi^{(1)} E_\omega(\vc{r},t) +
               \varepsilon_0 \chi^{(2)} E_\omega^2(\vc{r},t) + \ldots$,
where $\varepsilon_0$ is the vacuum permittivity, and $\chi^{(1)}$,
$\chi^{(p)}$, $(p=2,3,\ldots)$ are, respectively, the linear and the nonlinear
optical susceptibilities of order $p$.
Each one of the nonlinear polarization terms can be correlated to a harmonic
field $E^p(\vc{r},t)$, of order $p$, with phase and amplitude that are in
general different to that of the fundamental.
Phase memory in this case would be expressed as a preservation of the spatial
phase of the fundamental in the harmonics, and can be extremely beneficial when
using engineered optical wave packets~\cite{Glushko1993,Olivier2008,Liu2016} to
tailor the harmonic generation process.
For example, since the spatial phase controls the wave packet propagation, phase
memory would lead harmonics sharing the same propagation properties as the
fundamental.

Here we show that phase memory, during harmonic generation, is an inherent
property of a family of optical wave packets whose distribution is described by
the Airy function~\cite{Vallee2004}.
This family includes the accelerating Airy beams~\cite{Polynkin2009a,
Siviloglou2007a,Abdollahpour2010}, as well as the cylindrically symmetric
ring-Airy wave packets~\cite{Efremidis2010a,Papazoglou2011}.
The spatial phase distribution of Airy beams leads to unique properties,
including the propagation along parabolic trajectories~\cite{Siviloglou2007,
Polynkin2009a,Papazoglou2010}, bypassing obstacles and
self-healing~\cite{Broky2008,Baumgartl2008}.
Also, ring-Airy beams exhibit abrupt autofocusing, while experiencing only a
minor nonlinear focus shift as their power is
increased~\cite{Panagiotopoulos2013}, making them ideal candidates for high
intensity applications with precise deposition of energy in
space~\cite{Manousidaki2016,Liu2016}.
Beside these exiting properties, we demonstrate that phase memory preserves the
abruptly autofocusing behavior in the harmonics of ring-Airy wave packets, while
under certain conditions their foci coincide in space with the one of the
fundamental.
We analytically show that this behavior is further preserved under the action of
a converging lens for obtaining tighter focusing and higher intensities.
Experiments and detailed numerical simulations of the second harmonic of
ring-Airy beams validate our theoretical analysis.
Our results open the way for the use of accelerating beams and their harmonics
in a plethora of nonlinear optics applications.

\begin{figure}[t] \centering
  \includegraphics[width=0.45\textwidth]{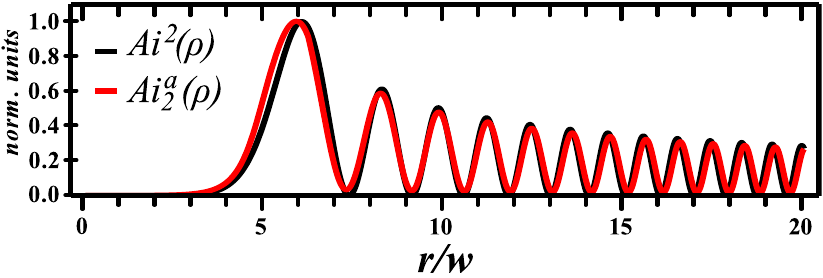}
  \caption{\label{fig1}%
           Approximation of the $\Ai^2$ function}
\end{figure}

{\it Theoretical analysis.}---%
Since the generation of harmonics involves powers of the field distribution we
will start our analysis form the case of second harmonic generation.
The amplitude of a ring-Airy beam is described as:
\begin{equation} \label{Ring_Airy_Function}
    u_\omega(r) = u_0 \Ai(\rho) e^{\alpha\rho},
\end{equation}
where $u_0$ is the amplitude, $\Ai(\cdot)$ is the Airy function,
$\rho\equiv(r_0-r)/w$, and $r_0$, $w$, and $\alpha$ are, respectively, the
primary ring radius, width and apodization parameters.
Without loss of generality the amplitude of the generated 2$^\mathrm{nd}$
harmonic in a thin BBO crystal is given by:
\begin{equation} \label{2d_Harmonic_Amplitude}
    u_{2\omega}(r) \propto \chi^{(2)} u_\omega(r)^2
                   = \chi^{(2)} u_0^2 \, \Ai(\rho)^2 \, e^{2\alpha\rho},
\end{equation}
where $\chi^{(2)}$ is the nonlinear susceptibility.
By using well-known approximations of the Airy function~\cite{Vallee2004} we
reach to an analytical expression of the square of the Airy function term that
appears in Eq.~\eqref{2d_Harmonic_Amplitude}~\cite{Supplement}:
\begin{equation} \label{Ring_Airy2_Aprox}
	\Ai(\rho)^2 \simeq 2E(\rho)^2 + E(\rho) \Ai(\hat{S}\cdot\rho)
                \equiv \Ai_2^a(\rho),
\end{equation}
where $E(\rho) \equiv 1/[2\sqrt{\pi}f(\rho)^{1/4}]$, $\hat{S}$ is a linear
scaling operator $\hat{S}\cdot\rho \equiv 2^{2/3}\rho + \pi/(8\times2^{1/3})$
and $f$ is an apodization function~\cite{Supplement}.
Figure~\ref{fig1} shows a comparison $\Ai(\rho)^2$ to the approximation
$\Ai_2^a(\rho)$.

Using the approximation of Eq.~\eqref{Ring_Airy2_Aprox} we can write for the
second harmonic:
\begin{equation} \label{2ndHarmonic}
	u_{2\omega}(r,0) \propto 2\chi^{(2)} u_0^2 E(\rho)^2 +
                            \chi^{(2)} u_0^2 \Ai(\rho') E(\rho) e^{2\alpha\rho},
\end{equation}
where $\rho' \equiv (r'_0-r)/w'$, $r'_0 = r_0 + \pi w/16$, $w' = 2^{-2/3} w$.
Equation~\eqref{2ndHarmonic} shows clearly that the amplitude of the generated
second harmonic is composed by two terms, a smooth pedestal that does not
exhibit any autofocusing behavior and a ring-Airy term that is strongly
apodized, carrying $\sim25$\% of the total energy.
The second term is a clear demonstration of phase memory, which will be
resulting in abrupt autofocusing of the second harmonic.

One can calculate the focus position of the fundamental, using the 1D Airy
analytical formulas~\cite{Siviloglou2007a,Papazoglou2011,Panagiotopoulos2013}:
\begin{equation} \label{Fai_omega}
	f_\Ai^\omega = 4\pi \frac{w^2}{\lambda} \sqrt{r_0/w + 1}
                 \simeq 4\pi \frac{w^{3/2}}{\lambda} \sqrt{r_0},
                 \quad r_0 \gg w.
\end{equation}

Under the same assumption ($r_0 \gg w$) we can estimate the autofocus position
of the second harmonic autofocusing term by using its primary ring radius $r_0'$
and width $w'$ parameters as estimated by Eq.~\eqref{2ndHarmonic}:
\begin{equation} \label{Fai_2omega}
    f_\Ai^{2\omega} = 4\pi \frac{w'^2}{\lambda/2} \sqrt{r_0'/w' + 1}
                    \simeq 4\pi \frac{w^{3/2}}{\lambda} \sqrt{r_0}
                    = f_\Ai^\omega.
\end{equation}
Interestingly, this result shows that the autofocusing component of the second
harmonic will autofocus at the same position as the fundamental.
So, the phase memory in the second harmonic acts is such a way that its
propagation dynamics are identical to the fundamental.
As for the pedestal term, it does not interfere with this focus since it
diffracts out.

We can now generalize this analysis for the generation of ring-Airy harmonics of
any higher order.
As we demonstrate in the following, all harmonics exhibit phase memory and
inherit the propagation dynamics of the fundamental.
In more detail, for the case of even orders $\Ai(\rho)^{2m}$, $(m=1,2,\ldots)$
Eq.~\eqref{Ring_Airy2_Aprox} is generalized to~\cite{Abramowitz1970,Supplement}:
\begin{equation} \label{Generic_Airy_even}
    \Ai(\rho)^{2m} \simeq \Ai_2^a(\rho)^m
                   = \sum\limits_{n = 0}^m \frac{2^{m - n} m!}{n! (m - n)!}
                     E(\rho)^{2m-n} \Ai(\hat{S} \cdot \rho)^n.
\end{equation}
Furthermore, whenever $n=2^l$, $(l=0,1,\ldots)$ the corresponding power term
$\Ai(\hat{S}\cdot\rho)^n$ in Eq.~\eqref{Generic_Airy_even} will result to a sum
of ``net'' ring-Airy terms $\sum_{i=-1}^{l-1}(\cdots)\Ai(\hat{S}^{l-i}\cdot\rho)$,
where $\hat{S}^q\cdot\rho \equiv \alpha_q\rho + \beta_q$, and
$\alpha_q \equiv 2^{2q/3}$,
$\beta_q \equiv \pi/(8 \times 2^{1/3}) (1 - 2^{2q/3})/(1 - 2^{2/3})$.
Each of these harmonic ring-Airy terms will autofocus at
positions~\cite{Supplement}:
\begin{align} \label{Fai_Generic_Airy_even}
	f_{\Ai(q)}^{(2m\,\omega)} & = 4\pi
                                     \frac{w_q^2}{\lambda_{(2m\,\omega)}}
                                     \sqrt{\frac{r_{0q}}{w_q} + 1}
                                   \simeq m \, 2^{1-q} \, f_\Ai^\omega, \\
                               q & = 1,\ldots,\log_2m + 1, \notag
\end{align}
where $\lambda_{(2m\,\omega)}=\lambda/2m$ is the $2m^\mathrm{th}$ harmonic
wavelength, $r_{0q}=r_0+\alpha_q\beta_qw$, $w_q=w/\alpha_q$, and $f_\Ai^\omega$
refers to the autofocus position of the fundamental.
Interestingly, in all the even harmonics that are a power of 2 (i.e.
$2^\mathrm{nd}$, $4^\mathrm{th}$, $\ldots$) there will exist a term that will be
autofocusing at $f_\Ai^\omega$.
On the other hand, for odd orders $\Ai(\rho)^{2m+1}$, $(m=1,2,...)$
Eq.~\eqref{Ring_Airy2_Aprox} is generalized to:
\begin{align} \label{Generic_Airy_odd}
    \Ai(\rho)^{2m+1} & \simeq \Ai_2^a(\rho)^m \, \Ai(\rho) \notag \\
                     & = \sum\limits_{n = 0}^m
                         \frac{2^{m-n}m!}{n!(m-n)!}
                         E(\rho)^{2m-n} \Ai(\hat{S}\cdot\rho)^n Ai(\rho).
\end{align}
Clearly now there exists only one ``net'' ring Airy term (for $n=0$).
This harmonic ring-Airy term will be autofocusing at~\cite{Supplement}:
\begin{equation} \label{Fai_Generic_Airy_odd}
	f_\Ai^{(2m+1)\,\omega} = 4\pi \frac{w^2}{\lambda_{(2m+1)\,\omega}}
                                \sqrt{\frac{r_0}{w} + 1}
                              \simeq (2m + 1) f_\Ai^\omega.
\end{equation}

\begin{figure}[t] \centering
    \includegraphics[width=0.45\textwidth]{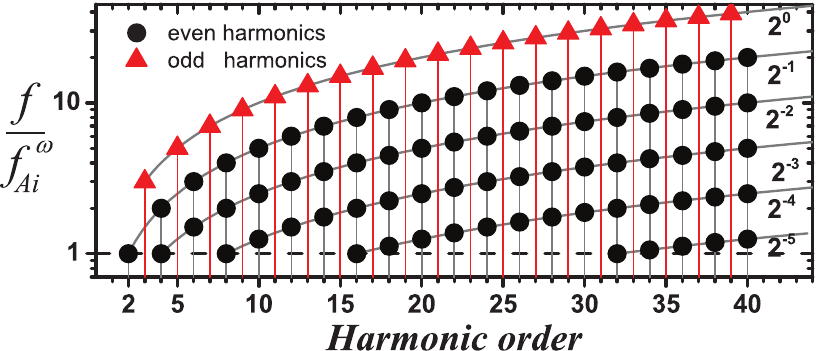}
    \caption{\label{fig2}%
             Autofocusing positions of ring-Airy harmonics.
             Note that even orders exhibit multiple foci (vertical lines are
             guides to the eye to identify each harmonic order).
             Curves that follow the points are $f = 2^j f_\Ai^\omega$,
             $(j=0, -1,\ldots)$, the factors $2^j$ are indicated on each curve.}
\end{figure}

Figure~\ref{fig2} shows an overview of these results up to the $40^\mathrm{th}$
harmonic.
Due to the phase memory, all harmonics exhibit abrupt autofocusing.
We can clearly see that even orders exhibit multiple foci, and that when the
order is a power of 2 there exists always a focus at $f_\Ai^\omega$.

\begin{figure}[t] \centering
  \includegraphics[width=0.45\textwidth]{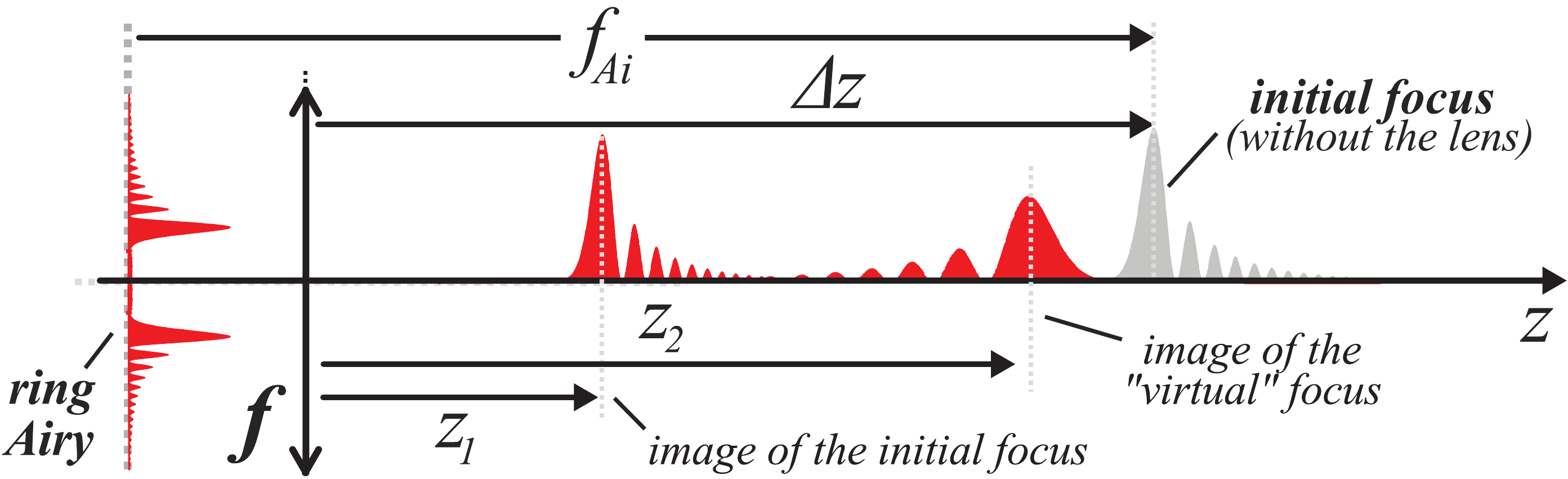}
  \caption{\label{fig3}%
           Focusing of a ring Airy beam by a thin lens.
           As in the case of all Janus waves two foci are
           observed~\cite{Papazoglou2016}.}
\end{figure}

The next challenge in our analysis is related to reaching higher intensities
compared to the ones achievable by autofocusing.
A straight forward way of doing this is by using focusing optical elements.
As we have previously shown~\cite{Papazoglou2016}, ring-Airy beams can be
further focused using a lens, though they behave in a peculiar way, they exhibit
double foci, a property characteristic to the family of Janus
waves~\cite{Papazoglou2010} where they belong.
The main question here is how their harmonics will be affected.
Although in the following we are discussing the effect that focusing has on the
second harmonic, our analysis can be easily expanded to the case of higher
harmonics as well.
As shown in Fig.~\ref{fig3} if we assume that a thin lens is located at a
distance $\Delta z$ from the focus of the ring-Airy beam we get for the
fundamental two foci at positions $z_1^\omega$, $z_2^\omega$ relative to the
lens~\cite{Papazoglou2016}:
\begin{equation} \label{Generic_Lens_Focus_Ai_omega}
	z_1^\omega = \left(\frac{\Delta\tilde{z}^\omega}
                            {\Delta\tilde{z}^\omega + \tilde{f}^\omega}
                 \right) f,
    \quad
    z_2^\omega = \left(\frac{\Delta\tilde{z}^\omega - 2}
                            {\Delta\tilde{z}^\omega - 2 + \tilde{f}^\omega}
                 \right) f,
\end{equation}
where, $\tilde{f}^\omega \equiv f/f_\Ai^\omega$,
$\Delta\tilde{z}^\omega \equiv \Delta z/f_\Ai^\omega$.
In the case of the second harmonic the situation is more complex.
The second harmonic Airy term will lead to two foci at positions
$z_1^{2\omega}$, $z_2^{2\omega}$ while the smooth pedestal term will focus at
position $z_3^{2\omega}$:
\begin{align} \label{Generic_Lens_Focus_Ai_2omega}
    z_1^{2\omega} & = \left(
                      \frac{\Delta\tilde{z}^{2\omega}}
                           {\Delta\tilde{z}^{2\omega} + \tilde{f}^{2\omega}}
                      \right) f
                    \simeq z_1^\omega, \notag \\
    z_2^{2\omega} & = \left(
                      \frac{\Delta\tilde{z}^{2\omega} - 2}
                           {\Delta\tilde{z}^{2\omega} - 2 + \tilde{f}^{2\omega}}
                      \right) f
                    \simeq z_2^\omega, \\
    z_3^{2\omega} & \cong f, \notag
\end{align}
where $\tilde{f}^{2\omega} \equiv f/f_\Ai^{2\omega} \simeq \tilde{f}^\omega$,
$\Delta\tilde z^{2\omega} \equiv \Delta z/f_\Ai^{2\omega}
                          \simeq \Delta\tilde{z}^{2\omega}$.
Interestingly, even after the focusing, the two foci of the second harmonic Airy
term overlap with those of the fundamental.
So, focusing does not deteriorate its phase memory inherited propagation
dynamics.

Note that our analysis applies also to the harmonics and powers of 1D and 2D
Airy beams, but in this case the resulting Airy terms in the harmonics do not
result in any distinguishable accelerating features, since their propagation is
shadowed by the propagation of the non-Airy terms~\cite{Dolev2010,Dolev2012,Dolev2012a}.
Likewise, the above can be applied in the temporal domain for Airy
pulses~\cite{Ament2011,Saari2009,Piksarv2014,Abdollahpour2010}.

\begin{figure}[t] \centering
  \includegraphics[width=0.45\textwidth]{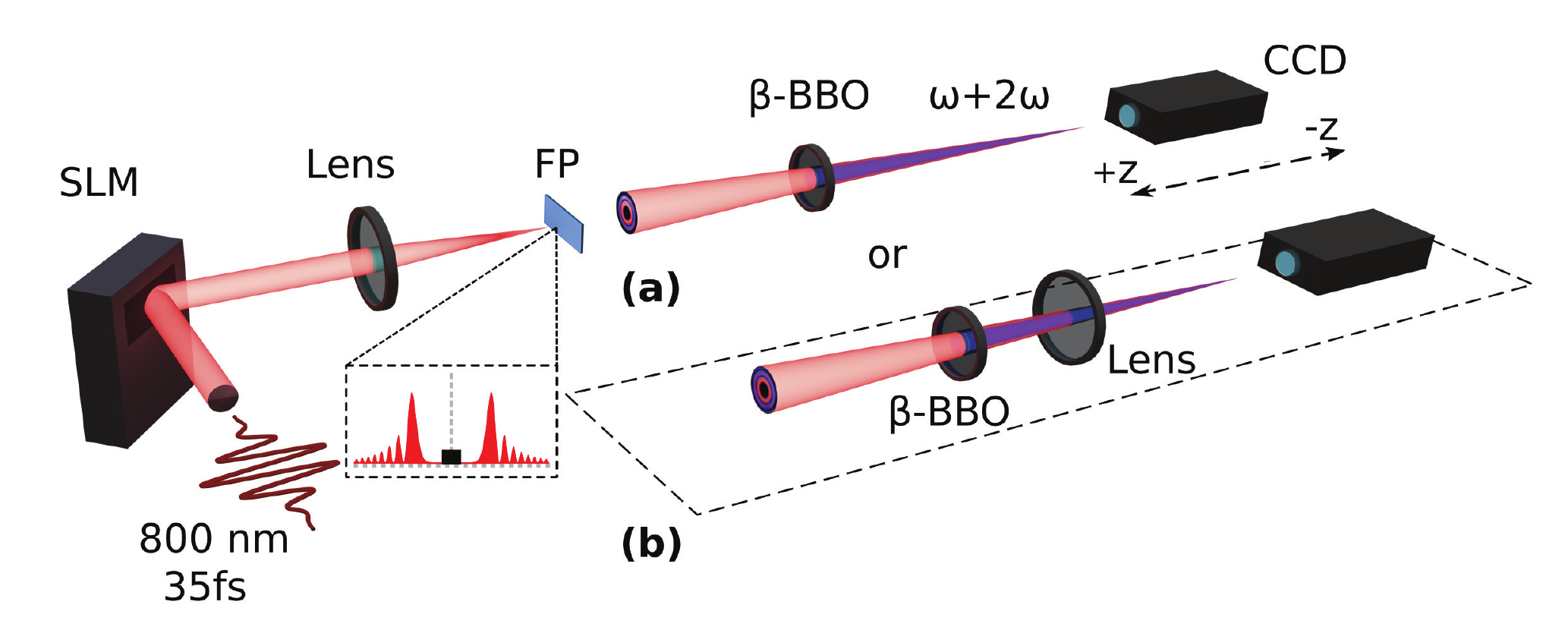}
  \caption{\label{fig4}%
           Schematic representation of the experimental setup.
           The ring Airy beam is generated at the Fourier plane (FP) of a lens
           after being phase modulated by a spatial light modulator (SLM).
           A beta-BBO generates the second harmonic.
           The fundamental and its second harmonic either propagate freely and
           autofocus (a) or are focused by a lens (b).
           The transverse profile of the beam is captured by a CCD camera at
           different distances along the propagation axis $z$.}
\end{figure}

{\it Experimental part.}---%
We have performed second harmonic (SH) generation experiments using autofocusing
ring-Airy beams.
In our experiments we studied the propagation of the fundamental and SH, both in
free propagation and after focusing by a lens.
The experimental setup used for the generation of the SH is shown in
Fig.~\ref{fig4}.
A Ti:Sapphire laser system, delivering Gaussian shaped beams at 800\,nm, 35\,fs
at 50\,Hz repetition rate, was used.
The ring-Airy beam was generated using a Fourier transform
approach~\cite{Papazoglou2011} in which the phase of the Gaussian laser beam was
modulated using a spatial light modulator (SLM, Hamamatsu LCOS-X10468-2).
The beam parameters (radius, width) were selected so that the beam abruptly
autofocuses at $f_\Ai^\omega=400$\,mm from the generation plane.
For generating the second harmonic the autofocusing beam propagated in a type-I
$\beta$-barium borate (BBO) crystal of 200\,$\mu$m thickness.
The conversion efficiency in this case was measured to be 7.3\%.
The fundamental and its second harmonic were then allowed either to freely
propagate or were focused using a focusing lens $f=100$\,mm.
The distances between the BBO crystal and that of the focusing lens from the FP
were 60\,mm and 146\,mm respectively.
Bandpass interference filters were used to isolate the fundamental and the
second harmonic.

The transverse intensity distribution of the beam along its propagation was
imaged by a linear CCD camera (14 bit), which was moved along the propagation
axis $z$.

\begin{figure}[t] \centering
  \includegraphics[width=0.45\textwidth]{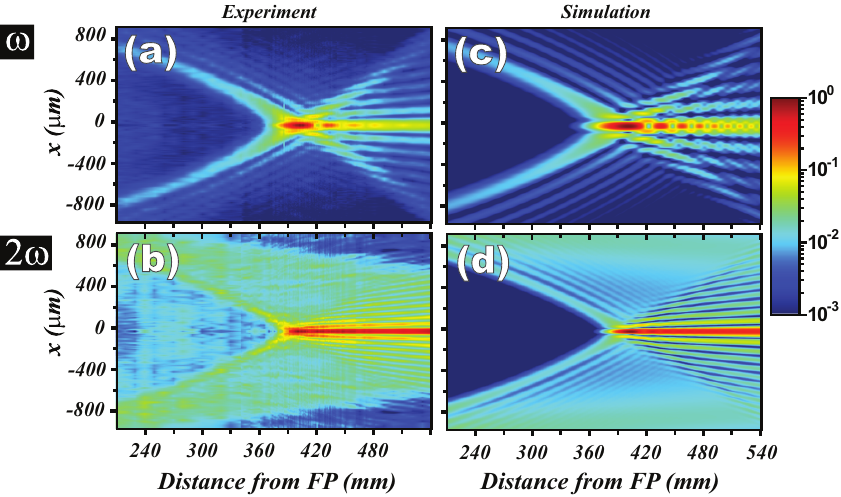}
  \caption{\label{fig5}%
           Free propagating fundamental and second harmonic ring Airy beams (a),
           (b) together with numerical simulations (c), (d).}
\end{figure}

In the first experiments we report below we studied the propagation and
autofocusing properties of the fundamental and its second harmonic without the
presence of any focusing lens.
Combining 2D transverse $(x,y)$ images that were captured at various positions
along the propagation $z$ axis, the $I(x,y,z)$ intensity profile of the beam was
retrieved.
An $x$-$z$ cross section of such a profile is shown in Fig.~\ref{fig5} for the
above case.
Figure~\ref{fig5}(a), reveals the autofocusing behavior of the fundamental
ring-Airy beam, which focuses at $z=400$\,mm as expected, followed by the
characteristic ring-Airy focus intensity distribution.
As predicted by our model and shown in Fig.~\ref{fig5}(b), the second harmonic
exhibits a parabolic trajectory, autofocusing at the same focus position as the
fundamental.
The non-ring-Airy pedestal term diffracts out and appears as a lower intensity
halo in the images.

Our analysis is completed through detailed numerical simulations of the SH
generation process and the propagation dynamics of the fundamental and its
second harmonic.
Our model is based on two coupled equations for monochromatic fields of the
fundamental and its second harmonic~\cite{Boyd2003}.
In the coupling terms we assumed perfect phase matching and the second order
susceptibility, $\chi^{(2)}$, equal to
$4\times 10^{-12}$\,m/V~\cite{Sutherland2003}.
The corresponding numerical simulations, shown in Figs.~\ref{fig5}(c), (d), are
in very good agreement with the experimental results, while the analytical
predictions of the foci positions nicely agree as well.

\begin{figure}[t] \centering
  \includegraphics[width=0.45\textwidth]{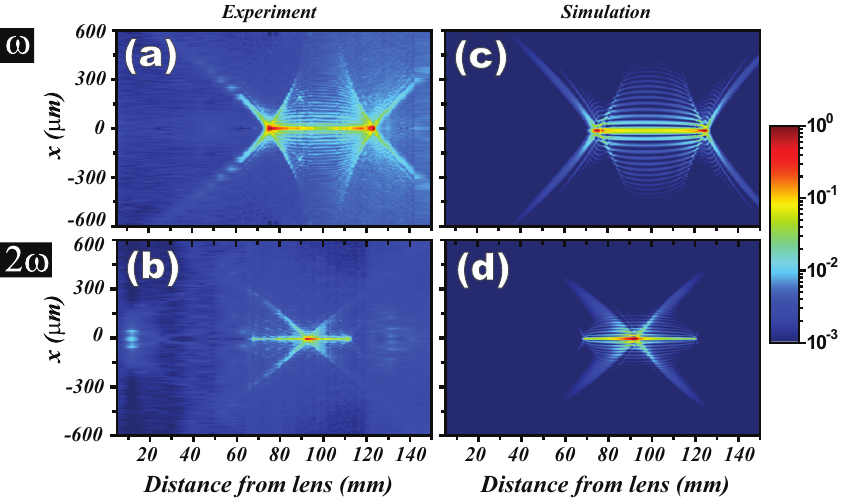}
  \caption{\label{fig6}%
           Fundamental and second harmonic ring-Airy beams after being focused
           by a lens.
           (a), (b) Experimental results  and (c), (d) numerical simulations.}
\end{figure}

Placing a focusing lens after the BBO crystal, results in the peculiar behavior
previously discussed.
Figures~\ref{fig6}(a), (c) show, respectively, the experimental and simulation
results for the fundamental beam, presenting two discrete conjugate foci
distributions.
Again, a very nice agreement is found between the results from our experiments,
simulations and analytical predictions, with the two foci located at
$z_1^\omega=75.3$\,mm and $z_2^\omega=122.6$\,mm.
Figures~\ref{fig6}(b), (d) show, respectively, the experimental and simulation
results for the second harmonic beam.
Once more, the agreement between the results from our experiments, simulations
and analytical predictions is nice.
The central intense focus results from the pedestal term and is located at
$z_3^{2\omega}=92.7$\,mm, practically at the focal plane of the lens.
The other two foci are positioned before and after the focal plane of the lens
at $z_1^{2\omega}=67.4$\,mm and $z_2^{2\omega}=112.8$\,mm, respectively.
The combination of the three foci results in an elongated focal volume for the
second harmonic, which, as one can clearly see in Fig.~\ref{fig6}, overlaps to a
great extend with the also elongated focal volume of the fundamental.
This extended spatial overlap of the fundamental and second harmonic is of
pivotal role in nonlinear wave-mixing experiments, like in the generation of
intense THz fields using two-color ring-Airy beams~\cite{Liu2016}.

{\it Conclusions.}---%
In conclusion, we have demonstrated theoretically and experimentally that the
harmonics of autofocusing ring-Airy beams preserve the phase of the fundamental.
Through an analytic approximation of the powers of the Airy function we have
shown that phase memory during harmonic generation is an inherent property of
all optical wave packets described by an Airy distribution.
Our findings are thus applicable to accelerating Airy beams (1D and 2D),
cylindrically symmetric ring-Airy beams but also in the temporal domain for Airy
pulses, and spatiotemporal Airy light bullets.
The phase memory in the case of ring-Airy beams results in abruptly autofocusing
harmonics, with their focus position coinciding with that of the fundamental for
even harmonics of power 2.
We have also demonstrated that even after focusing these beams still spatially
overlap, surprisingly over elongated focal volumes.
Our analytical predictions are in excellent agreement with second harmonic
generation experiments and detailed numerical simulations.
Our results open the way for using accelerating beams and their harmonics in a
plethora of nonlinear optics applications, like for intense THz
fields~\cite{Liu2016,Moradi2015b}, nonlinear wave-mixing~\cite{Dolev2010} and
filamentation~\cite{Polynkin2009a,Panagiotopoulos2013,Couairon2007b}.

\begin{acknowledgements}
This research was supported by the National Priorities Research Program grant
No.~NPRP9-383-1-083 from the Qatar National Research Fund (member of The Qatar
Foundation), the H2020 "Laserlab Europe" (EC-GA 654148) and the H2020 "MIR-BOSE"
(EC-GA 737017).
\end{acknowledgements}

\bibliographystyle{apsrev4-1}
\bibliography{PhaseMemoryHarmonics.bib}

\end{document}